# Myths and Realities About Online Forums in Open Source Software Development: An Empirical Study

Faheem Ahmed[1,*], Piers Campbell[1], Ahmad Jaffar[1] and Luiz Fernando Capretz[2]

[1]College of Information Technology, United Arab Emirates University, P. O. Box 17551, Al Ain, Abu Dhabi, United Arab Emirates

[2]Department of Electrical & Computer Engineering, University of Western Ontario, London, Ontario, Canada N6A 5B9

**Abstract:** The use of free and open source software (OSS) is gaining momentum due to the ever increasing availability and use of the Internet. Organizations are also now adopting open source software, despite some reservations, in particular regarding the provision and availability of support. Some of the biggest concerns about free and open source software are post release software defects and their rectification, management of dynamic requirements and support to the users. A common belief is that there is no appropriate support available for this class of software. A contradictory argument is that due to the active involvement of Internet users in online forums, there is in fact a large resource available that communicates and manages the provision of support. The research model of this empirical investigation examines the evidence available to assess whether this commonly held belief is based on facts given the current developments in OSS or simply a myth, which has developed around OSS development. We analyzed a dataset consisting of 1880 open source software projects covering a broad range of categories in this investigation. The results show that online forums play a significant role in managing software defects, implementation of new requirements and providing support to the users in open source software and have become a major source of assistance in maintenance of the open source projects.

## I. INTRODUCTION

In the recent past many large software development companies have committed their efforts to open source projects which gave momentum to this initiative. The term OSS refers to software equipped with licenses that provide existing and future users the right to use, inspect, modify, and distribute (modified and unmodified) versions of the software to others [1]. Open source is a software development methodology that makes source code available to a large community that participates in development by following flexible processes and communicating via the Internet [2]. The favorable acceptance of OSS products by business and the direct involvement of major IT vendors in OSS development have transformed OSS from a fringe activity, developed for public good, to a mainstream, commercially viable form of software development [3]. As an economically viable alternative to closed source software, OSS has been proven to reduce maintenance costs, a fact that benefits all stakeholders, except profit driven vendors. Active open source projects usually have a well-defined community with common interests that are involved in either continuously evolving the product (or related products) or in using its results [4]. OSS is developed by loosely organized communities of participants located around the world and working over the Internet. Remarkably, most participants contribute without being employed, paid, or recruited by the organization [5]. The use of Internet further accelerates the popularity and use of free and open source software at an unprecedented rate of growth.

The process of software maintenance in OSS is different from the traditional method of software development. For example, change management in a traditional setting is generally required to have a change control authority present in the organization to approve or reject the new features or changes in requirements. Approved modifications are typically then assigned to a predefined set of individuals for implementation. The decision of approval or rejection is heavily based on the impact of the changes on the overall software application. This scenario is very different to most OSS cases, where it is not mandatory to get the change request approved from any authority. In OSS anyone can propose a change and perform the implementation by themselves or a request can be floated among participants in the community. In some cases there are moderators present in the OSS project communities who supervise the activities of the OSS projects, and who are required to evaluate the change request before approving any implementation. Although a recent trend is for companies that benefit from OSS to have some employees work in the OSS domain, surveys have shown that the majority of such participants are volunteers [6]. The Floss Survey [7] identified many other reasons why developers were involved in OSS development, including becoming part of the open source community, promoting the open source mode of development, supporting the idea of "free" as an alternative to proprietary software, gaining a reputation, and having fun.

*Address correspondence to this author at the College of Information Technology, United Arab Emirates University, UAE; Tel: 00971-3-7135521; Fax: 00971-3-762-6309; E-mail: f.ahmed@uaeu.ac.ae

## A. Research Motivations

It is not only the concept of providing "free" access to the software and its source code that makes OSS the phenomenon that it is, but also the development culture. The collaborative nature of the OSS culture makes use of a wide volunteer community which conducts its development activities in a decentralized environment that has the direct result of effectively lowering production costs and improving the software quality [8]. According to his book, "The cathedral and the bazaar", Raymond [8] draws an analogy between "The Cathedral" (Proprietary Software) and "The Bazaar" (OSS), where "the Cathedral" development is carefully crafted by individuals in an isolated work place. In this "cathedral" approach there are no beta releases as everything has been fixed to a single plan, single point of focus, or even single mind [9]. Conversely, "The Bazaar" style is similar to a communal meeting place, a setting in which parties add different elements to the interaction. It is within this community of like-minded individuals that discussions take place and ideas and information regarding the software are disseminated [8, 9]. The main mechanism for the exchange of ideas and collaboration in the OSS arena is the on-line forum. It is via these forums that community members discuss ideas related to their common projects, present evaluations, and suggest enhancements (such as new feature requests). The forums are also the main channel through which requests for support are made by users, who are not necessarily members of the development community, and the forums are also the medium for the reporting of defects in the software. This model of both community interaction and support provision is extremely dissimilar to the traditional model used in proprietary software development.

OSS however, is still perceived as risk-laden alternative to proprietary software as support is deemed to be of a higher level with the latter class of software. This is despite of the wide community of developers and clear indications from literature that OSS produces quality products which can offer significant advantages when compared to proprietary products. This perception is driven by the concerns over the availability of support which is available for a given OSS product. As identified the main recourse for users is via online forums and there is a commonly held belief that such a "voluntary" medium does not provide a reliable support service. The same views also exist around the areas of defect reporting and feature request. Therefore in this work we have set out to empirically examine the role of on-line user forums, the main point of interaction for users, and the critical areas of defect reporting, managing requests for support and feature requests. Our intention is to examine the evidence available to assess whether this commonly held belief is based on facts given the current developments in OSS or simply a myth which has developed around OSS development.

## B. Literature Survey

OSS is increasingly being acknowledged as a viable alternative to commercial proprietary software, with signifi-cant software reliability and value for money benefits for businesses. Proprietary software is closed source, available at a cost, and its copyright is retained by the developing organization. Consequently the end user does not have access to the source code, is unable to customize the software to their individual requirements, and the software cannot be redistributed. With OSS, however, the user has access to the source code, enabling them to customize the software to fit their needs and, if required, redistribute it [10]. It is important to note however that OSS is not exempt of copyright and there are some licensing schemes that protect contributions of the author(s) but which also permit the redistribution of the software and even allow for additional derivative development [11]. Industry is not alone in recognizing the opportunities presented by OSS, indeed a number of governments around the world [12] have acknowledged the significance of this kind of software and research into the viability, usability, maintainability, and supportability of OSS is growing significantly. A case in point is the European Union (EU), which has shown significant interest in OSS, in particular through its development of the e-Europe initiative [13], which encourages the adoption of OSS by member countries. This interest in OSS has been driven by the increased technical awareness of the general public and the necessity to provide faster more efficient services via technical channels. This public demand has lead to an e-transformation in governmental operations [14].

Empirical studies regarding open source quality assurance activities and quality claims are rare [15]. Koponen [16] discusses defect management and version management systems as an integral part of OSS maintenance process. Aberdour [17] observes that the open source software model has led to the creation of significant pieces of software, and many of these applications show levels of quality comparable to closed source software development. Raymond suggests the high quality of OSS can be achieved due to the high degree of peer review and user involvement in bug/defect detection. Generally a popular or active project means that the community in the OSS project are interacting constantly and providing feedback to activities such as defect identification, fixing of defects, new feature request and support requests for the further improvement. Wayner [18] found that developers contribute from around the world, meet face-to-face infrequently, if at all, and coordinate their activities primarily by means of computer-mediated communications. Crowston and Scozzi [19] investigated the coordination practices for software bug fixing in OSS development teams and observed that task sequences are mostly sequential and composed of few steps, namely; submit, fix and close and that effort is not equally distributed among process actors, indicating that few actors contribute heavily to all tasks, while the majority just submit one or two bugs. Cubranic and Booth [20] discussed major issues of coordinating open source development projects, including collaborative communication mediums and configuration management tools. Mockus *et al.* [21] provided a comprehensive comparison of Apache against five commercial products in terms of developer participation, team size, productivity and defect density, and problem resolution.

Open source software is developed by a community of likeminded individuals who freely provide access to the software source code. A clear benefit of providing the source code at the time of distribution is that end users are then able to learn more about the functionality and operation of the software program [22]. A clear implication of providing unrestricted access to the source code is that other interested parties are able to amend the source code to customize or improve the original and then make the amended version publicly available. This notion of freely available software does not therefore solely refer to a lack of purchasing cost, but also includes freedom of development, usage, and distribution [23]. In proprietary software project development, developers are typically motivated by money. However, at its most fundamental and pure level the OSS development concept is not motivated by financial reward but rather by what could be viewed as some alternative compensation, such as aiding society or demonstrating their individual proficiency. According to Lerner & Tirole [24] these two motivation types can be classified as immediate pay-off and delayed benefit. In the case of Immediate pay-off (proprietary software) a monetary incentive is the most common motivation for any software developments, while delayed benefits are typically realized as indirect economic benefits that programmers will get in the future [24]. Woods and Guliani [25], illustrated the OSS development cycle and showed that central to the OSS development process is the community of users and developers who both create and drive the process. As previously stated OSS projects are unique in their developed environment in as much as they consist of communities sharing a common interest create the software. These communities are also responsible for the creation of a control or governance structure which in turn controls the project lifecycle. This governance structure grows from individual motivation and develops into one social control mechanism [10, 26, 27]. This social control creates conformity for certain moral and cultural rules within the development community [10].

It is this special culture that has driven the popularity of OSS, firstly, within personal (individual) user circles and secondly and more recently, in corporate and governmental circles. It is the intention that in many cases the developers of OSS seek no financial reward yet produce a realistic alternative that has driven its popularity as an alternative to profit driven software vendors. Indeed Zeitlyn [9] compared the relationship of OSS developers and their projects to that of parents and their children, in as much as parents always give everything to their children without expecting anything in return. In addition to this reasoning a number of other benefits of OSS have been identified [25, 27, 28, 29, 30]. Acquisition cost of OSS is generally lower than proprietary software and may even be completely free of cost and thus may eliminate the financial burdens of proprietary licensing schemes.

The unique culture of OSS development is therefore not suited to the same styles of project management as the used in the proprietary industry. As a result two types of control have been identified by Latteman & Stieglitz [10] in OSS projects. The first, "direct governance" is a social control that ensures the quality of a project by employing more traditional techniques (as found in industry driven projects) such as conducting reviews and monitoring performance and progress of tasks associated to the project. The second approach

"indirect governance" is a less structured controlling process with a focus on assessing the quality of a project based on its output, in other words how good the completed project is and how well it is received by the community of users. This community based development approach of OSS enables software solutions to be fully customized according to the functional needs of the organization and to be tested on a global scale. However, proprietary software is designed according to vendor's development planning and follows a common design, which lacks the depth and breadth allowed by OSS [25]. In proprietary software, software quality testing is limited within a controlled environment and specific scenarios [24]. However, OSS development involves much more elaborate testing, as OSS solutions are tested in various environments, by various skillsets and experiences levels of programmers, and are tested in various geographic locations around the world [24, 31-33].

According to Kessler [34] the Internet provides planning and organizational resources as well as cost-effective communication and distribution systems. OSS projects can be characterized as communities of users where information, assistance and innovation are freely shared [35]. Some recent empirical studies found that the average OSS contributor is approximately 30 years old and highly educated [36, 37]. In OSS projects, online communication via the Internet is used to capture and store knowledge through the systematic collection and coding of tasks in files and lists, e.g. development lists, where discussion centers on topics pertaining to the next release of the software in question [38]. Scacchi [39] examined the process of requirements engineering in open source projects and provided a comparison with traditional processes. Scacchi [40] found that little is known about how people in these communities coordinate software development across different settings, or about what software processes, work practices, and organizational contexts are necessary to ensure their success.

## II. RESEARCH MODEL & HYPOTHESES OF THE STUDY

The world has witnessed a rapid growth of OSS development projects after the increased popularity and use of Internet. This has formed a diverse community of software developers all across the globe, who share directly or indirectly knowledge and communicate using online forums. Their needs and interests are also diverse. An indirect measure of success and failure of an OSS can be considered as the activities on the online forums. An online forum for OSS having a continuous increase in message volume shows the high level of interest and helps in identifying and fixing defects, managing new features and support requests. The research model shown in Fig. (**1**) deals with the association of online forums with software defects, new features and support requests. We further formulated three research questions RQ-1, RQ-2 and RQ-3 to summarize the research objectives of this study. The purpose of research question RQ-1 in this study is to analyze the association between software defect identification and fixing in OSS and public forums associated with the OSS project. Research question RQ-2 analyzes the association between managing dynamic requirements in the form of new feature requests in OSS and online public forums associated with the OSS project. The third objective of the research model explained in research question RQ-3 is to understand the association between OSS support and online public forums associated with the OSS project. In summary the main objectives of this study are to investigate the answer to the following three research questions:

**RQ-1:**     Do online public forums help in identifying and fixing OSS defects?

In order to empirically investigate the research question RQ-1 we hypothesize the following:

**H1:**     Online public forums help in identifying and fixing OSS defects.

H11:     The Open bugs present in OSS are positively related with mailing lists in online forums.

H12:     The open bugs in OSS are positively related with the number of messages in online forums.

H13:     The close bugs present in OSS are positively related with mailing lists in online forums.

H14:     The close bugs in OSS are positively related with the number of messages in online forums.

**RQ-2:**  Do online public forums help in managing dynamic requirements in OSS projects?

In order to empirically investigate the research question RQ-2 we hypothesize the following:

**H2:**     Online public forums help in managing dynamic requirements in OSS project.

H21:     The Open feature requests present in OSS are positively related with mailing lists in online forums.

H22:     The open feature requests in OSS are positively related with the number of messages in online forums.

H23:     The closed feature requests present in OSS are positively related with mailing lists in online forums.

H24:     The closed feature requests in OSS are positively related with the number of messages in online forums.

**RQ-3:**  Do online public forums help in managing the support activities in OSS projects?

In order to empirically investigate the research question RQ-3 we hypothesize the following:

**H3:**     Online public forums help in managing support activities in OSS project.

H31:     The Open support requests present in OSS are positively related with mailing lists in online forums.

H32:     The open support requests in OSS are positively related with the number of messages in online forums.

H33:     The close support requests present in OSS are positively related with mailing lists in online forums.

H34:     The close support requests in OSS are positively related with the number of messages in online forums.

It is important to mention at this juncture that we are using the term "open bug" as a defect, which is indentified but has not yet been fixed, whereas "close bug" refers to a bug which was reported and fixed. We are using the phrase "open feature requests" as a change in requirements to add more functionality that has not been accommodated yet, whereas "close feature requests" refers to a change in requirements which has been implemented. The term "open support requests" is any activity such as, installation help, documentation, information retrieval, usability, etc. and has not been accommodated yet, whereas "close support requests" refers to a one such activity which was implemented. Online forums are consisting of one or more "mailing lists" which constitute the discussion forum where "messages" are exchanged.

## III. DATA COLLECTION & EXPERIMENTAL SETUP

We collected data for 1880 open source software projects from www.sourceforge.net, a popular data repository for open source software projects on the Internet. The dataset covers various categories of open source software projects such as; communication, database, desktop, education, format & protocols, games & entertainment, scientific & engineering, security, software development, system and text editor. The first filtration activity removed the data concerning projects which had either a total of software defect, new feature or support requests equal to 0 or that had no online forums. The dataset is therefore reduced from 1880 projects to 650. Subsequently, outliers selected on the basis of total software defects, new feature or support requests and a number of online forums were removed and this further reduced the dataset to 616 open source projects. Fig. (**2**) illustrates the number of new feature requests in various open source software projects in the initial dataset of this study. It also highlights the outliers which were removed and the updated distribution of total new feature requests is shown in Fig. (**3**). Figs. (**4 & 5**) illustrate the distribution of total online forums versus number of messages in the forums in various open source software projects before (Fig. **4**) and after (Fig. **5**) removing the outliers. Fig. (**4**) illustrates that most of the number of messages in open source software falls from 0 to 4000. Whereas some projects have higher than this range and some project have even 30,000 messages. In fact the number of projects having messages more than 4000 are very few in comparison to the number of messages less than 4000 and truly not representing the population therefore we removed these projects and in Fig. (**5**) we presents the data after removing these projects and it clearly shows that the population of the dataset used in the study falls in the range of 0 to 4000 number of messages in the public forums.

In the dataset of this study we used communication (183), database (76), desktop (48), education (21), format & protocols (15), games & entertainment (60), scientific & engineering (41), security (47), software development (28), system (53) and text editor (44) projects. Fig. (**6**) illustrates the distribution of the dataset in these categories. The maximum open bugs were found in the category of format & protocols (215). The minimum number of (67) open bugs was observed in the category of system software. The maximum number of bugs which have been fixed were observed in the categories of database and desktop environment (857 each).

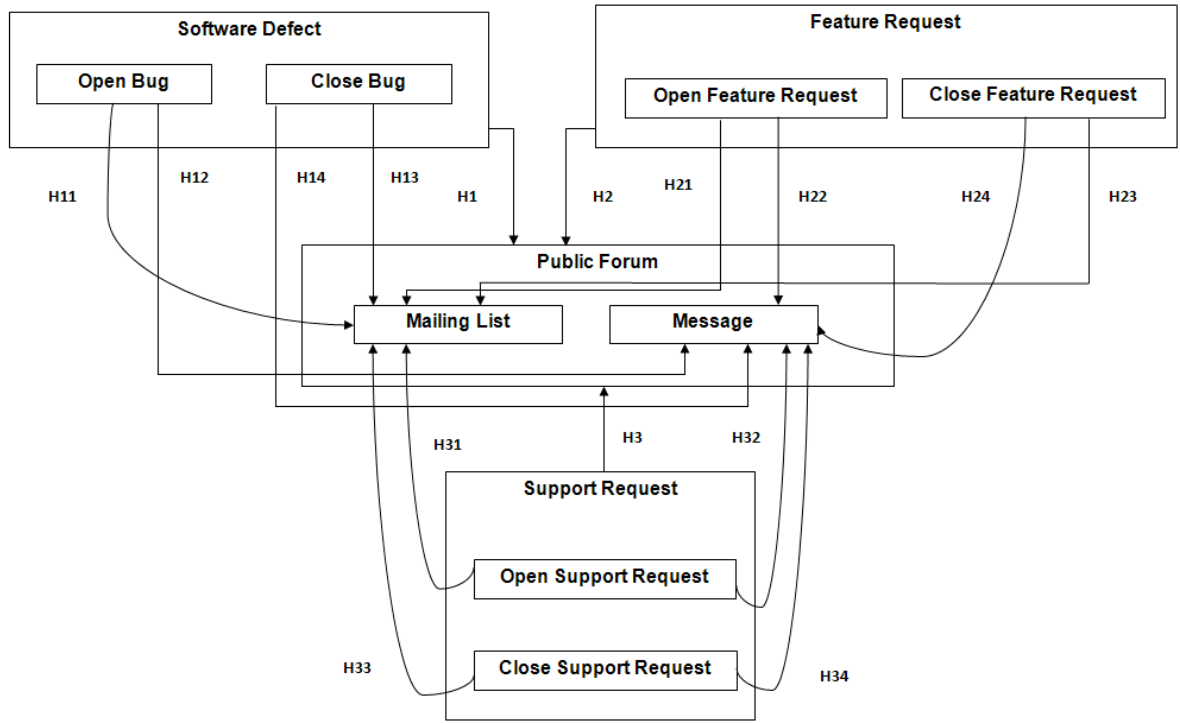

**Fig. (1).** Research model of the study.

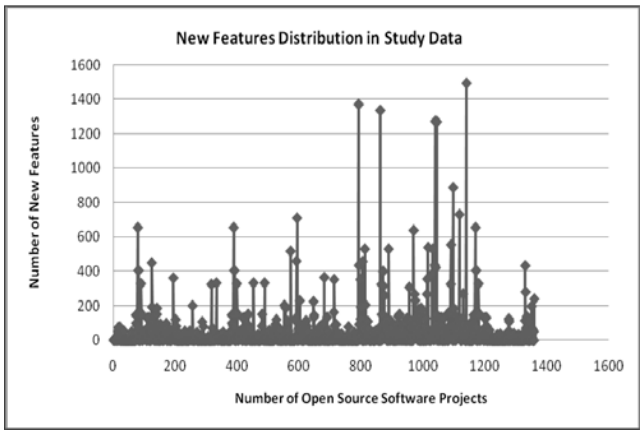

**Fig. (2).** New features distribution of study dataset.

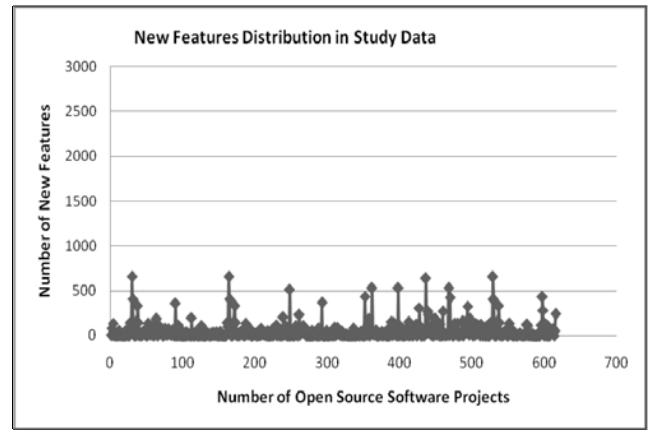

**Fig. (3).** New features distribution after removing outliers.

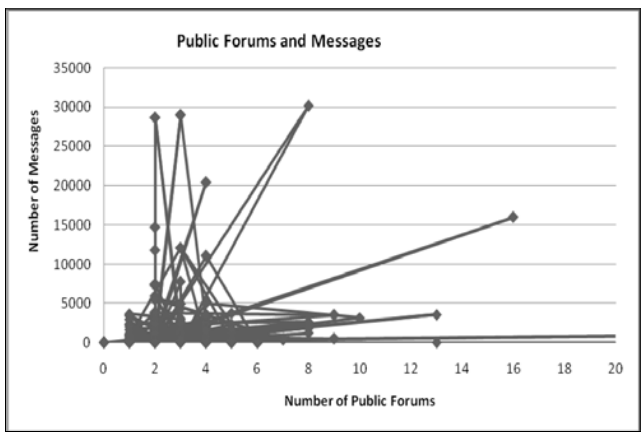

**Fig. (4).** Public forums and number of messages in forums data distribution.

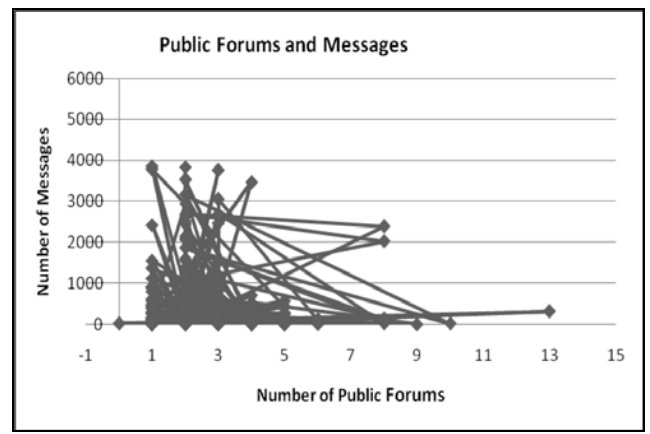

**Fig. (5).** Public forums and number of messages in forums data distribution after removing outliers.

The category of system software also shared the minimum number of (344) bugs which have been fixed. Database, desktop environment and format & protocol categories had observed maximum number (868 each) of total bugs. The category of software development project has minimum number of (507) known total bugs. The category of "communication" has maximum number of online forums (13) in one project. The highest occurrence of open feature requests was found in the category of communication, games/entertainment and scientific and engineering (276). The minimum number of (50) open feature requests were observed in the category of software development. The maximum for feature requests which have been implemented was observed in the categories of database and desktop environment (432 each).

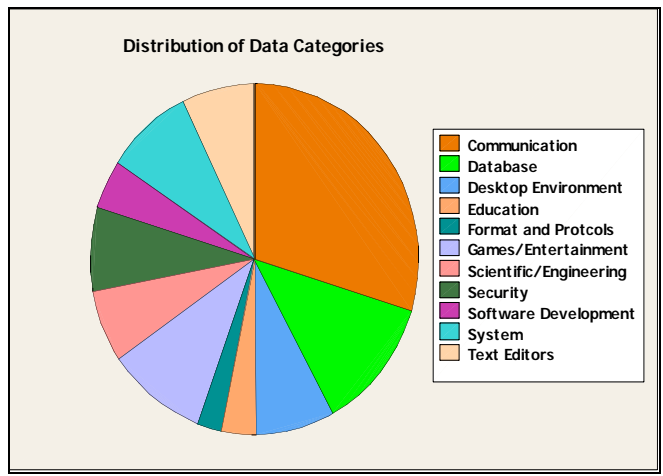

**Fig. (6).** Distribution of project categories in the dataset.

The category of system software also shared the minimum number of (100) feature requests that have been implemented. Communication, games/entertainment and scientific and engineering had the observed maximum number (655 each) of total feature requests. The category of software development project had minimum number of (146) known total feature requests. The category of "communication" has the maximum number of online forums (13) in one project. The highest number of messages (5611) was found in the category communication projects. The lowest number of (2938) messages was observed in the category database projects. The maximum number of mailing lists of (7) each was observed in the categories communication, education, games & entertainment and scientific & engineering projects. Desktop environment, security and text editors shared the minimum number of (5) mailing lists in a project.

### A. Intuitive Understanding of Dataset

In order to provide intuitive understanding of the dataset used in this study, we are presenting in this section the graphical relationship of various entities involved in the research hypothesis. A regression line shown in each diagram further enhanced the understanding of the dataset and their relationship. All the diagrams are generated using Minitab 1.5 statistical tool. Fig. (**7**) illustrate the relationship of open and close bugs with open and close support requests. An "open bug" means that a bug has been reported but has not been fixed yet, whereas "close bugs" refers to one which was reported and fixed. On the other hand "open support request" means that a request to support has been placed but have not yet been served, whereas "close support request" refers to one that has been requested and served. It is intuitively seen from Fig. (**7**) more support requests are present if there are more open bugs and whereas if more bugs are closed and fixed it also closed more support requests as well.

Fig. (**8**) illustrate the relationship of open and close bugs with open and close new feature requests. Again in this figure we are freeing "open bug" as a bug that has been reported but has not been fixed yet, whereas "close bugs" refers to one which was reported and fixed. On the other hand "open new feature request" means that a request to add a new feature has been placed but have not yet been completed, whereas "close new feature request" refers to one that

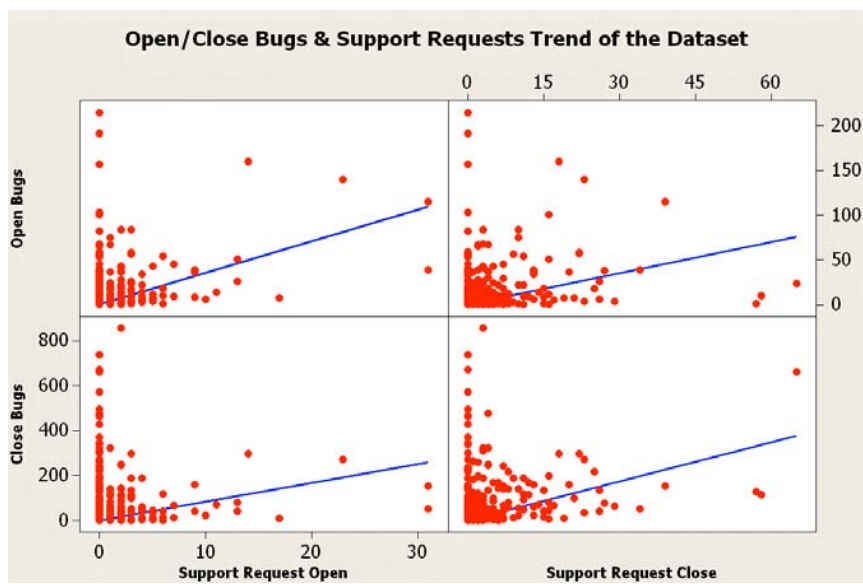

**Fig. (7).** Open/Close bugs & support requests trend in the dataset.

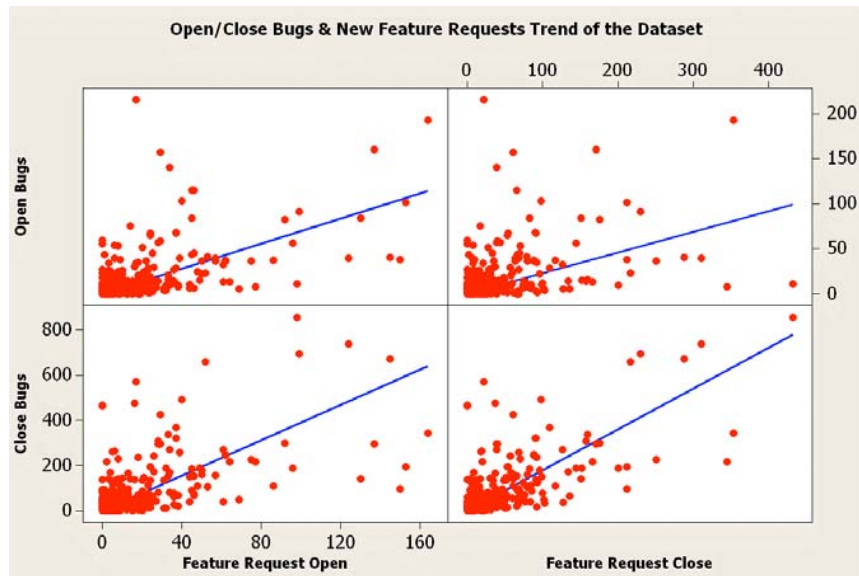

**Fig. (8).** Open/Close bugs & new feature requests trend in the dataset.

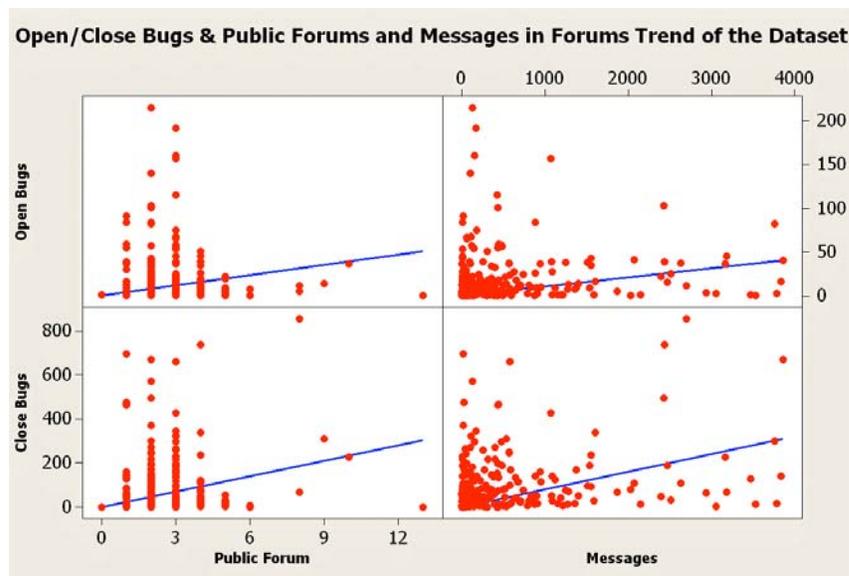

**Fig. (9).** Open/Close bugs & public forums and messages in forums trend in the dataset.

has been requested to add a new feature and it is completed as well. It is intuitively seen from Fig. **(8)** more new feature requests are closed if there are more bugs that has been fixed and whereas if more bugs are yet open and has not been fixed than we have more new feature requests that has not been served yet. Fig. **(9)** illustrate the relationship of open and close bugs with number of public forums and number of messages in public forums. Number of forums refers to the mailing lists where users communicate messages. Whereas message simulates either a new initiative for discussion or reply to an existing threads. It is intuitively seen from Fig. **(9)** that if we have more public forums then that results in more closed bugs in comparison to open bugs. Similarly, in case when there are more messages in the public forums than it results in more bugs that have been fixed.

## IV. DATA ANALYSIS & RESULTS

To analyze the research model and check the significance of hypotheses H1, H2, and H3 and their sub-hypotheses we used various statistical analysis techniques. Initially we divided the data analysis activity into three phases. Phase-I dealt with normal distribution tests and parametric statistical analysis. Phase-II dealt with non-parametric statistical analysis. In order to increase the external validity of the study, we used both statistical approaches of parametric and non-parametric methods. We tested for the normal distribution of all factors of total, open, close bugs, total new feature requests, open and close new feature requests, total support request, close and open support requests as well as number of online forums, mailing list and messages using mean, standard deviation, kurtosis and skewness techniques, and found the values for all these tests to be within the acceptable range for the normal distribution with some minor exceptions.

**Table 1. Data Analysis Results of Research Question (RQ-1)**

| | Software Defects | | | | | |
|---|---|---|---|---|---|---|
| | **Structural Tests** | | **Pearson Correlation** | | **Spearman Correlation** | |
| | **Open Defects** | **Close Defect** | **Open Defects** | **Close Defect** | **Open Defects** | **Close Defect** |
| Mailing List | (H14) Coefficient: 0.17 $R^2$: 0.03 F-ratio: 19.69 | (H13) Coefficient: 0.22 $R^2$: 0.05 F-ratio: 34.07 | (H14) Coefficient: 0.43 P< 0.001 | (H13) Coefficient: 0.22 P< 0.001 | (H14) Coefficient: 0.80 P< 0.01 | (H13) Coefficient: 0.58 P< 0.01 |
| Messages | (H12) Coefficient: 0.09 $R^2$: 0.08 F-ratio: 57.71 | (H11) Coefficient: 0.43 $R^2$: 0.19 F-ratio: 146.46 | (H12) Coefficient: 0.29 P< 0.001 | H11) Coefficient: 0.17 P< 0.001 | (H12) Coefficient: 0.71 P< 0.01 | H11) Coefficient: 0.73 P< 0.01 |

We conducted tests for hypotheses H1, H2, and H3 and their sub-hypotheses using parametric statistics, such as the Pearson correlation coefficient and one tailed t–test in Phase-I. In Phase-II of non-parametric statistics, we conducted tests for hypotheses using the Spearman correlation coefficient. Phase-III dealt with testing the hypotheses of the research model of this study using the Partial Least Square (PLS) technique. The PLS technique helps when complexity, non-normal distribution, low theoretical information, and small sample size are issues [41, 42]. In the PLS testing of hypotheses we keep one factor as independent and other as dependent variable. We used the PLS technique to increase the reliability of the results. The statistical calculations were performed using Minitab® 14 software.

We examined the Pearson correlation coefficient and t-test between variables involved in the hypotheses H11, H12, H13 and H14 which deal with software defects and online forums in OSS. The Pearson correlation coefficient between open bugs and number of mailing lists in the public forums was positive (0.17) at P < 0.001, and thus provided a justification to accept the H11 hypothesis. The hypothesis H12 was accepted based on the Pearson correlation coefficient (0.29) at P < 0.001, between open bugs and number of messages on the online forum. The correlation coefficient of 0.22 at P < 0.001 was observed between the close bugs and number of mailing lists in the online forum. The positive correlation coefficient of 0.43 at P < 0.001 meant that H14 was accepted. Hence, it was observed and is reported here that hypotheses H11, H12, H13, and H14, were found statistically significant and were accepted.

In Phase-II we conducted non-parametric statistical technique using Spearman correlation coefficient to test the hypotheses H11, H12, H13 and H14. Hypothesis H11 was statistically significant at P < 0.01 with Spearman correlation coefficient of 0.73. A positive association was observed between open bugs and number of messages (H12) on the online forum (Spearman: 0.71 at P < 0.01). H13, which deals with between the close bugs and number of mailing lists in the online forum, was accepted (Spearman: 0.58 at P < 0.01.

The Spearman correlation of (0.80 at P < 0.01) was observed for H14. Hence, it was observed and is reported here that hypotheses H11, H12, H13, and H14, were found statistically significant and were accepted. In Phase-III of hypotheses testing, we used the PLS technique to overcome some of the associated limitations and to cross validate with the results observed using the approaches of Phase-I and Phase-II.

We tested the hypothesized relationships, i.e. H11, H112, H13 and H14, by examining their direction and significance. The hypotheses involved two variables therefore in PLS we placed one variable as the response variable and other as the predicate. It contains observed values of path coefficient, $R^2$ and F-ratio. The path coefficient open defect (H11) was found to be 0.17, $R^2$: 0.03 and F-ratio (19.69) was significant at P < 0.001. Open defects (H12) had positive path coefficient of 0.09 with $R^2$: 0.08 and at P < 0.001 F-ratio was 57.71 with number of messages. Close defect with mailing list (H13) (Path coefficient: 0.22, $R^2$: 0.05, F-ratio: 34.07 at P < 0.001) had the same direction as proposed. Close defect and number of messages (H14) (Path coefficient: 0.43, $R^2$: 0.19, F-ratio: 146.46 at P < 0.001) also had the same direction as proposed in H2b. All in all, the hypotheses H11, H12, H13 and H14, showed significant at P < 0.001 with a positive path coefficient and were in the same direction as proposed, therefore illustrates that the hypothesis H1 is accepted and hence concluded that online public forums help in identifying and fixing OSS defects, which provides answer to the RQ-1. Table **1** reports the results of the structural tests, Pearson correlation and Spearman correlation of the hypotheses.

We examined the Pearson correlation coefficient and t-test between variables involved in the hypotheses H21, H22, H23 and H24, which deal with new feature requests and online forums in OSS. The Pearson correlation coefficient between open feature requests and number of mailing lists in the public forums was positive (0.12) at P < 0.005, and thus provided a justification to accept the H21 hypothesis. The hypothesis H22 was accepted based on the Pearson correlation coefficient (0.55) at P < 0.001, between open feature requests and number of messages on the online forum. The correlation coefficient of 0.16 at P < 0.001 was observed between the close feature requests and number of mailing lists in the online forum. The positive correlation coefficient of 0.51 at $P < 0.001$ meant that H24 was accepted. Hence, it was observed and is reported here that hypotheses H21, H22, H23, and H24, were found statistically significant and were accepted.

**Table 2. Data Analysis Results of Research Question (RQ-2)**

| | Feature Requests | | | | | |
|---|---|---|---|---|---|---|
| | Structural Tests | | Pearson Correlation | | Spearman Correlation | |
| | Open Feature Requests | Close Feature Requests | Open Feature Requests | Close Feature Requests | Open Feature Requests | Close Feature Requests |
| Mailing List | (H24)<br>Coefficient: 3.2<br>$R^2$: 0.12<br>F-ratio: 9.9 | (H23)<br>Coefficient: 9.11<br>$R^2$: 0.02<br>F-ratio: 17.19 | (H24)<br>Coefficient: 0.51<br>$P < 0.001$ | (H23)<br>Coefficient: 0.16<br>$P < 0.001$ | (H24)<br>Coefficient: 0.42<br>$P < 0.001$ | (H23)<br>Coefficient: 0.23<br>$P < 0.001$ |
| Messages | (H22)<br>Coefficient: 0.02<br>$R^2$: 0.30<br>F-ratio: 269.79 | (H21)<br>Coefficient: 0.04<br>$R^2$: 0.26<br>F-ratio: 218.11 | (H22)<br>Coefficient: 0.55<br>$P < 0.001$ | (H21)<br>Coefficient: 0.12<br>$P < 0.005$ | (H22)<br>Coefficient: 0.41<br>$P < 0.001$ | (H21)<br>Coefficient: 0.22<br>$P < 0.001$ |

In Phase-II hypothesis H21 was statistically significant at $P < 0.001$ with Spearman correlation coefficient of 0.22. A positive association was observed between open feature requests and number of messages (H22) on the online forum (Spearman: 0.41 at $P < 0.001$). H23, which deals with between the close feature requests and number of mailing lists in the online forum, was accepted (Spearman: 0.23 at $P < 0.001$. The Spearman correlation of (0.42 at $P < 0.001$) was observed for H24. Hence, it was observed and is reported here that hypotheses H21, H22, H23, and H24, were found statistically significant and were accepted.

We tested the hypothesized relationships, i.e. H21, H22, H23 and H24, by examining their direction and significance. It contains observed values of path coefficient, $R^2$ and F-ratio. The path coefficient open feature requests (H21) was found to be 3.2, $R^2$: 0.12 and F-ratio (9.9) was significant at $P < 0.001$. Open feature requests (H22) had positive path coefficient of 0.02 with $R^2$: 0.30 and at $P < 0.001$ F-ratio was 269.79 with number of messages. Close feature requests with mailing list (H23) (Path coefficient: 9.11, $R^2$: 0.02, F-ratio: 17.19 at $P < 0.001$) had the same direction as proposed. Close feature requests and number of messages (H24) (Path coefficient: 0.04, $R^2$: 0.26, F-ratio: 218.11 at $P < 0.001$) also had the same direction as proposed in H2b. All in all, the hypotheses H21, H22, H23 and H24, showed significant at $P < 0.001$ with a positive path coefficient and were in the same direction as proposed, therefore illustrates that the hypothesis H2 is accepted and hence concluded that online public forums help in managing dynamic requirements in OSS project, which provides answer to the RQ-2. Table **2** reports the results of the structural tests, Pearson correlation and Spearman correlation of the hypotheses.

We examined the Pearson correlation coefficient and t-test between variables involved in the hypotheses H31, H32, H33 and H34. The Pearson correlation coefficient between open support requests and number of mailing lists in the public forums was positive (0.08) at $P < 0.005$, and thus pro-vided a justification to accept the H31 hypothesis. The hypothesis H32 was accepted based on the Pearson correlation coefficient (0.25) at $P < 0.01$, between open support requests and number of messages on the online forum. The correlation coefficient of 0.20 at $P < 0.01$ was observed between the close support requests and number of mailing lists in the online forum. The positive correlation coefficient of 0.23 at $P < 0.01$ meant that H34 was accepted. Hence, it was observed and is reported here that hypotheses H31, H32, H33, and H34, were found statistically significant and were accepted.

In Phase-II hypothesis H31 was statistically significant at $P < 0.001$ with Spearman correlation coefficient of 0.32. A positive association was observed between open support requests and number of messages (H32) on the online forum (Spearman: 0.38 at $P < 0.01$). H33, which deals with between the close support requests and number of mailing lists in the online forum, was accepted (Spearman: 0.29 at $P < 0.01$. The Spearman correlation of (0.45 at $P < 0.01$) was observed for H34. Hence, it was observed and is reported here that hypotheses H31, H32, H33, and H34, were found statistically significant and were accepted.

We tested the hypothesized relationships, i.e. H31, H32, H33 and H34, by examining their direction and significance. It contains observed values of path coefficient, $R^2$ and F-ratio. The path coefficient open support requests (H31) was found to be 1.93, $R^2$: 0.80 and F-ratio (4.59) was significant at $P < 0.03$. Open support requests (H32) had positive path coefficient of 1.01 with $R^2$: 0.06 and at $P < 0.001$ F-ratio was 43.08 with number of messages. Close support requests with mailing list (H33) (Path coefficient: 4.01, $R^2$: 0.03, F-ratio: 2.07 at $P < 0.001$) had the same direction as proposed. Close support requests and number of messages (H34) (Path coefficient: 1.02, $R^2$: 0.20, F-ratio: 154.2 at $P < 0.001$) also had the same direction as proposed in H34. All in all, the hypotheses H31, H32, H33 and H34, showed significant at $P < 0.001$ with a positive path coefficient and were in the same direction as proposed, therefore illustrates that the hypothesis H3 is accepted and hence concluded that online public forums help in managing support activities in OSS project, which provides answer to the RQ-3. Table **3** reports the re-sults of the structural tests, Pearson correlation and Spearman correlation of the hypotheses.

**Table 3. Data Analysis Results of Research Question (RQ-3)**

| | Support Requests | | | | | |
|---|---|---|---|---|---|---|
| | **Structural Tests** | | **Pearson Correlation** | | **Spearman Correlation** | |
| | **Open Support Requests** | **Close Support Requests** | **Open Support Requests** | **Close Support Requests** | **Open Support Requests** | **Close Support Requests** |
| Mailing List | (H34) Coefficient: 1.93 $R^2$: 0.80 F-ratio: 4.59 | (H33) Coefficient: 4.01 $R^2$: 0.03 F-ratio: 2.07 | (H34) Coefficient: 0.23 $P< 0.01$ | (H33) Coefficient: 0.20 $P< 0.01$ | (H34) Coefficient: 0.45 $P< 0.01$ | (H33) Coefficient: 0.29 $P< 0.01$ |
| Messages | (H32) Coefficient: 1.01 $R^2$: 0.06 F-ratio: 43.08 | (H31) Coefficient: 1.02 $R^2$: 0.20 F-ratio: 154.2 | (H32) Coefficient: :0.25 $P< 0.01$ | (H31) Coefficient: 0.08 $P< 0.005$ | (H32) Coefficient: 0.38 $P< 0.01$ | (H31) Coefficient: 0.32 $P< 0.001$ |

## V. DISCUSSION OF RESULTS & THREATS TO EXTERNAL VALIDITY

Open source software is increasing in popularity due to the volume of involvement in its management and use. One of the major reasons behind this advancement is the economic aspects and the other notable reason is the use of the Internet, which has virtually scaled down the world into a knowledge village. It is clear from our analysis that there is a positive correlation between the volume of messages posted on online forums and the number of open bugs reported in a particular OSS project. This demonstrates that the OSS community is active in testing projects and the identification of defects. It further highlights that defects are not simply accepted as an unavoidable feature of OSS but rather, the community, which is established around an OSS, project work collaboratively to identify and correct defects in a given project. Our study also shows that in addition to the support network generated around an OSS project, the volume of interested parties is significantly increased for projects with unsolved defects. This is demonstrated by the positive correlation between the number of open bugs (defects) and number of individuals who are registered in the mailing lists of a given project. This correlation suggests that the OSS community has a significant support network which is likely to be larger than the support team of a proprietary application. As mailing list members are also altruistic in nature it is likely that a collaborative environment will lead to a number of possible solutions being identified.

Further evidence of this collaborative support network that facilitates the fixing of defects in OSS can be witnessed in our analysis in the correlation between the number of messages in the online forums and the volume of fixed bugs. As can clearly be seen this positive correlation shows that the more active a thread on a particular OSS project is the greater the number of defects which have been closed. The same correlation can be observed when examining the volume of users in the mailing list and the number of closed bugs. Once again a highly active mailing list is positively correlated to the number of defects which have been rectified. The analysis clearly demonstrates that the interest and high level of involvement in OSS by volunteer developers leads to a high degree of available support which in turn leads to a rapid identification and subsequent rectification of defects in the projects.

Online forums play a highly important role in the whole development cycle of the OSS, most notably, in their success and failure when considered from the viewpoint of the live activities present in the project. These activities involve the exchange of messages in the online forum, and messages can be of different types ranging from support requests, bug identification and fixing, new feature request etc. As our analysis has shown there is a clear positive relationship between the number of open features and the number of messages in the online forum. The OSS community is renowned for its close interaction between professional and amateur software developers and the development character of OSS ensures that reuse is a central pillar in project development. Typically therefore a member of the OSS community will request features with the expectation that other developers have already created similar or exact matches which can be integrated into the product.

OSS development is primarily user driven and the motivation of community members, such as Sourceforge.net, is generally the continual enhancement of software, particularly when the application in question provides an open alternative to proprietary applications. As a result the interest in such applications is high and the support base drives forward the development of new features which are subsequently included in the successive releases. The primary conduits for such requests are the forum(s) related to the software, however, in order to interact with a forum it is required that users are registered in the mailing list associated with the software application. Consequently and as verified by our study, the greater the number of registered mailing list users the higher the number of open features. The OSS community is highly active, as clearly shown by the number of projects included in this study and we have already shown that OSS development and enhancement is correlated to the interaction with mailing list and forums. Our study also identifies a positive correlation between both the number of forum messages and size of mailing list and the number of closed features relating to each application. We propose that this correlation is also due to the spirit of reuse and collegiality engendered through the community of open source developers. Feature requests received from the forums and mailing lists are then developed, refined and incorporated into the application, which results in the closure of the of the request.

The role of support is also significantly important in the OSS development cycle, most notably, the success and failure of an OSS is typically dependent on the perception and actual availability of support for users. Normally, support is provided through the exchange of messages in the online forum and it is clear from our analysis that the number of open support requests and the number of online messages are positively related. The OSS community is renowned for its close interaction of professional and amateur software developers and the development characteristics of OSS projects ensure that reuse is a central pillar in project development. As such the user community will request support in the hope that other developers have already encountered the same issues and can provide solutions or recommendations regarding the product. Support requests will generally arrive via a common channel, namely the forums or mailing lists, but requests can cover a diverse range of areas such as documentation, installation issues, malfunctions during operation and many others.

The development style of OSS is very much user driven and it is this distinctive style and its associated community which enables this varied range of support requests to be submitted via a common channel, while distinct support channels would normally be used to support commercial applications. In providing a single point of contact which is able to provide feedback on support issues and typically a number of responses, OSS has established a support network in its own community which enables OSS to be considered a viable alternative to proprietary software. Once again the number of available OSS projects identifies the high level of activity in the community and OSS in general. Our study identified a positive correlation between both the number of support requests and size of mailing list and the number of closed support requests relating to each application. Again we propose that this correlation is due to the spirit of reuse and collegiality engendered through the open source community. As support is requested via forum messages (and mailing lists) by registered participants in the community, they are quickly serviced by other members, resulting in the closure of the support request. It is this ongoing cycle of requesting and solving (which may include developing and implementing) which makes open source software such an appealing area of software development.

### A. Threats to External Validity

Surveys, experiments, metrics, case studies, and field studies are examples of empirical methods used to investigate both software engineering processes and products [43]. Threats to external validity are conditions that limit the researcher's ability to generalize the results of his/her experiment to industrial practice [44], which was the case with this study. Specific measures were taken to support external va-lidity, for example, a random sampling technique was used to draw samples from the population in order to conduct experiments, and filtering was applied to the data set in order to remove the outliers. We retrieved the data from the most active and well-known OSS reporting website which has huge volume of projects listed. We sampled the data from various categories of software development mode, environment, programming languages, and hardware and software platform to generalize the study. The increased popularity of the empirical methodology in software engineering has also raised concerns regarding ethical issues [45, 46]. We followed the recommended ethical principles to ensure that the empirical investigation conducted and reported here would not violate any form of recommended experimental ethics. Therefore the data repository we used in this study is a non-profit organization. Another aspect of validity is concerned with whether or not the study reports results that correspond to previous findings. This study reinforces the current perceptions that the OSS movement is gaining momentum and that the OSS development life cycle is heavily dependent on online forums and their activities. Although the proposed approach has some potential to threaten external validity, we followed appropriate research procedures by conducting and reporting tests to improve the reliability and validity of the study, and certain measures were also taken to ensure the external validity.

### CONCLUSION

Free and open source software is gaining popularity at an unprecedented rate of growth. Organizations, despite some concerns about quality, have been using and are opting to use this type of software. The commonly believed myths concerning OSS state that online forums play a significant role in managing and development of OSS projects and their success and failure is heavily dependent on the live activities on the forums. The objective of this study was to analyze empirically the association between managing software defects, dynamic requirements and support in OSS and online public forums associated with the OSS project, thus finding realities of the established myths. We observed that online forums are the corner stone of managing software defects, dynamic requirements and support activities in OSS. The management of software defects, new feature or support requests right from inception, through elaboration to implementation is communicated via online forums. This study further helps in understanding the significant role of online forums in OSS development. The results of this empirical study provide evidence about the realities of some myths related to online forums in OSS projects.

---